\documentclass[final,5p,times,twocolumn]{elsarticle}

 \usepackage{svg}
 \usepackage{amsmath}
 \usepackage{tabularx}
 \usepackage{xparse,regexpatch}
 \usepackage{comment}
 \usepackage{hyperref}

\usepackage{amssymb}
\usepackage{lipsum}

\usepackage{lineno}

\newcommand\ddfrac[2]{\frac{\displaystyle #1}{\displaystyle #2}}
\NewDocumentCommand{\n}{m m}{\ensuremath{\mathrm{^{#1}#2}}}

\journal{ }

\begin{document}

\begin{frontmatter}



\title{High-precision beam profile measurement with a microchannel-plate detector in the high magnetic field of the WISArD experiment}


\author[LP2iB]{S. Lecanuet}
\author[LPC]{X. Fléchard}
\author[LP2iB]{P. Alfaurt}
\author[LP2iB]{P. Ascher}
\author[SCK]{D. Atanasov}
\author[LP2iB]{B. Blank}
\author[LP2iB]{L. Daudin}
\author[LPC]{H. DePreaumont}
\author[LP2iB]{M. Gerbaux}
\author[LP2iB]{J. Giovinazzo}
\author[LP2iB]{S. Grévy}
\author[LP2iB]{G. Guignard}
\author[IBS,KULeuven]{J. Ha}
\author[KULeuven]{C. Knapen}
\author[LP2iB]{S. Lechner}
\author[LP2iB]{A. Lépine}
\author[LPC]{J. Lory}
\author[LPC]{J. Perronnel}
\author[LP2iB]{M. Pomorski}
\author[LP2iB]{M. Roche}
\author[LP2iB]{C. Roumegou}
\author[KULeuven]{N. Severijns}
\author[IBS]{Y. Son}
\author[KULeuven]{S. Vanlangendonck}
\author[LP2iB]{M. Versteegen}
\author[UJF]{D. Zakoucky}

\affiliation[LP2iB]{organization={Université de Bordeaux, CNRS/IN2P3, LP2i Bordeaux UMR5797},
            postcode={F-33170}, 
            city={Gradignan},
            country={France}}

\affiliation[LPC]{organization={Université de Caen Normandie, ENSICAEN, CNRS/IN2P3, LPC Caen UMR6534},
            postcode={F-14000}, 
            city={Caen},
            country={France}}

\affiliation[IBS]{organization={Center for Exotic Nuclear Studies, Institute for Basic Science},
            postcode={34126}, 
            city={Daejeon},
            country={Republic of Korea}}
            
\affiliation[KULeuven]{organization={KU Leuven, Instituut voor Kern- en Stralingsfysica},
            postcode={B-3001}, 
            city={Leuven},
            country={Belgium}}

\affiliation[UJF]{organization={Nuclear Physics Institute, Acad. Sci.},
            postcode={CZ-25068}, 
            city={Rez},
            country={Czech Republic}}

\affiliation[SCK]{organization={Belgian Nuclear Research Centre SCK-CEN}, 
            postcode={Boeretang 200, 2400}, 
            city={Mol},
            country={Belgium}}

\begin{abstract}
We present the development and characterization of a compact low-energy ion beam diagnostic for the WISArD (Weak Interaction Studies with \n{32}{Ar} Decay) experiment at ISOLDE/CERN. The microchannel plate (MCP) detector, which is configured in a Z-stack and has a resistive position sensitive anode, was tested with both stable and radioactive beams. This work focuses on the image reconstruction method, which corrects the pincushion distortion inherent to the square-shaped resistive anode, and investigates the influence of the magnetic field on the detector performance. Our results demonstrate that the detector achieves beam profile measurements with sub-millimeter accuracy, while coping with the spatial and high-magnetic field (4 T) constraints of the experiment. These capabilities meet the precision requirements of the WISArD experiment for extracting the modified beta-neutrino angular correlation coefficient, $\tilde{a}_{\beta\nu}$, with an uncertainty of 1\textperthousand.
\end{abstract}



\end{frontmatter}


\section{Introduction}
\label{introduction}
\begin{figure}[h]
    \centering
    \includegraphics[width=1.\linewidth]{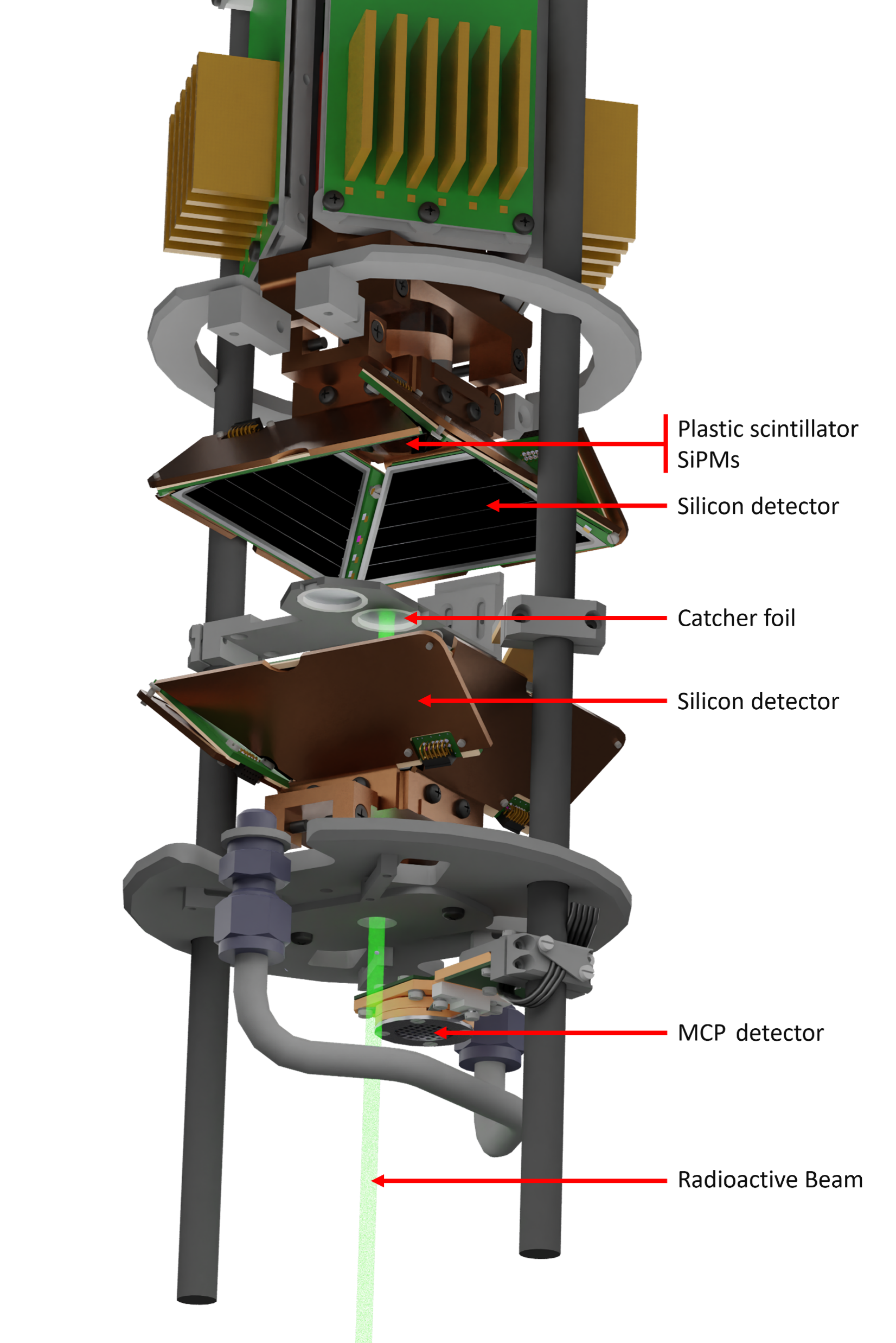}
    \caption{CAD view of the WISArD detection setup with the radioactive beam in fluorescent green implanted on the catcher. The MCP has been integrated at the bottom of the tower (here in \textit{out} position). See Ref.~\cite{Dinko_2023} for further details.}
    \label{FIG_tower}
\end{figure}

Microchannel plate (MCP) detectors are widely used in ato\-mic, nuclear, high-energy and anti-matter physics to detect single particles such as photons, electrons, and ions. When coupled to a position sensitive anode, these detectors provide both the detection time and the spatial coordinates of particles, achieving typical accuracies better than 100~ps and 100~µm for time and position, with detection efficiencies larger than 50\% for ions \cite{Lienard_2005}. Such capabilities make MCP detectors particularly suitable for precise measurements of the full momentum of low-energy particles \cite{Ullrich_1997,Muller_2022} and for beam imaging applications~\cite{COECK_2006,Baldin_2014,WIGGINS_2017}.
In the WISArD (Weak Interaction Studies with \n{32}{Ar} Decay) experiment at ISOLDE/CERN, which probes the electroweak interaction through high-precision measurements in nuclear beta decay, knowledge of the radioactive beam profile in the experimental setup is crucial. The goal of the experiment is the determination of the modified beta-neutrino angular correlation coefficient, $\tilde{a}_{\beta\nu}$ \cite{Gonzalez_2016}, in the beta decay of \n{32}{Ar}, with a precision of 1\textperthousand~ in order to probe exotic current beyond the Standard Model \cite{Falkowski_2021}. It relies on the measurement of the energy shift of the delayed protons emitted after beta decay, as a function of the relative angle between these protons and the beta particles~\cite{Victoria_2020}.
In the Monte-Carlo simulations used to extract $\tilde{a}_{\beta\nu}$, this relative angle critically depends on the location of the decay source. The ISOLDE facility delivers a 30~keV radioactive ion beam of \n{32}{Ar}, which is implanted into a thin aluminized mylar catcher foil at the center of the detection system.
The analysis of the WISArD proof-of-principle experiment revealed a systematic error of $\Delta \tilde{a}_{\beta\nu} = 4$\textperthousand~ resulting from the 3~mm uncertainty estimated on both the position and the radius of the implantation region~\cite{Victoria_2020}. To address this significant source of systematic error, a high-precision beam diagnostic with sub-millimeter accuracy was developed for integration into the upgraded experimental setup \cite{Dinko_2023}. The WISArD detection tower, shown in Fig.~\ref{FIG_tower}, is inserted in the former WITCH \cite{Finlay_2016} superconducting magnet which provides the high magnetic field (4~T) required for the experiment but offers limited space for such a diagnostic. We refer the reader to Ref.~\cite{Dinko_2023} for a detailed description of the WISArD detection setup. Given the above mentioned constraints, a compact MCP detector coupled to a square resistive anode was built and tested. \\
This article describes the detector design (Sec.~\ref{SEC_Mecanical_design}), details the image reconstruction and the detector’s performance (Sec.~\ref{SEC_Experimental_results})
Finally, Sec.\ref{SEC_Beam_Profile} presents the beam profile measurements obtained during the 2025 run at ISOLDE with radioactive \n{32}{Ar} ions, and discuss their impact on the extraction of the beta-neutrino angular correlation coefficient $\tilde{a}_{\beta\nu}$.

\section{MCP detector design}
\label{SEC_Mecanical_design}
\subsection{Choice of the position sensitive anode and MCP assembly}
The design of the beam monitor is constrained by two main factors: the high magnetic field and the limited space within the WISArD tower. 
An MCP-based detector was selected as it is well-suited for detecting ions with typical incident energies of several keV, while their compact size is advantageous for integration into the WISArD setup. However, it is well established that strong magnetic fields significantly reduce the gain of MCPs. This effect is due to the shortening of electron trajectories between two impacts on the microchannel surface leading to the reduction of secondary emission within the microchannels \cite{Li_2020}. Typically, a magnetic field of the order of 1~T can reduce the gain of standard 25~µm channel MCPs by one order of magnitude.
By using a smaller channel diameter and bias angle \cite{Lehmann_2008} one can mitigate significant gain loss from operating the detector in an environment of 4~T (targeted field for the online measurements). As a compromise between costs and performances, the F1551-01 from Hamamatsu, with a 17.9~mm plate diameter, 12~µm channel diameter and  8° bias angle was chosen. Three MCPs were mounted in a Z-stack configuration, which provides higher gain than the standard chevron configuration and increases the total charge emitted by the MCP assembly.
For position-sensitive readout, several anode types are commonly used with MCPs, including: delay lines, phosphor screens, backgammon anodes, and resistive anodes. The strong magnetic field of WISArD confines the electron cloud emitted by the back MCP to the scale of the microchannels, making anodes requiring significant electron cloud spreading (such as backgammon or delay lines) unsuitable. Phosphor screens, which require a CCD camera and optical coupling, were incompatible with the available space. 
Resistive anodes \cite{Gear_1969,Lampton_1979} are compact and compatible with a high magnetic field but commercially available detectors based on the Gear-anode geometry \cite{Gear_1969} were too large for the WISArD tower with a maximum diameter of 12~cm. Instead, a custom-made square-shaped resistive anode was fabricated. The resistive layer was created by mixing graphite powder with glyceride-based paint at a mass ratio of 1.00 (paint) to 1.33 (graphite), yielding a homogeneous mixture with suitable viscosity. This mixture was applied to a 200~µm deep, 16~mm × 16~mm square pocket milled into a PEEK\footnote{Polyether ether ketone} substrate, which serves as the anode holder. Excess paint was removed with a flat ruler to ensure a uniform 200~µm thickness. After drying, the per-square resistance R$_\square$ of the anodes that were produced typically ranged between 1~kΩ and 2~kΩ.

\subsection{Mechanical design}
\begin{figure}[ht]
    \begin{minipage}[l]{0.49\linewidth}
        \includegraphics[width=1\linewidth]{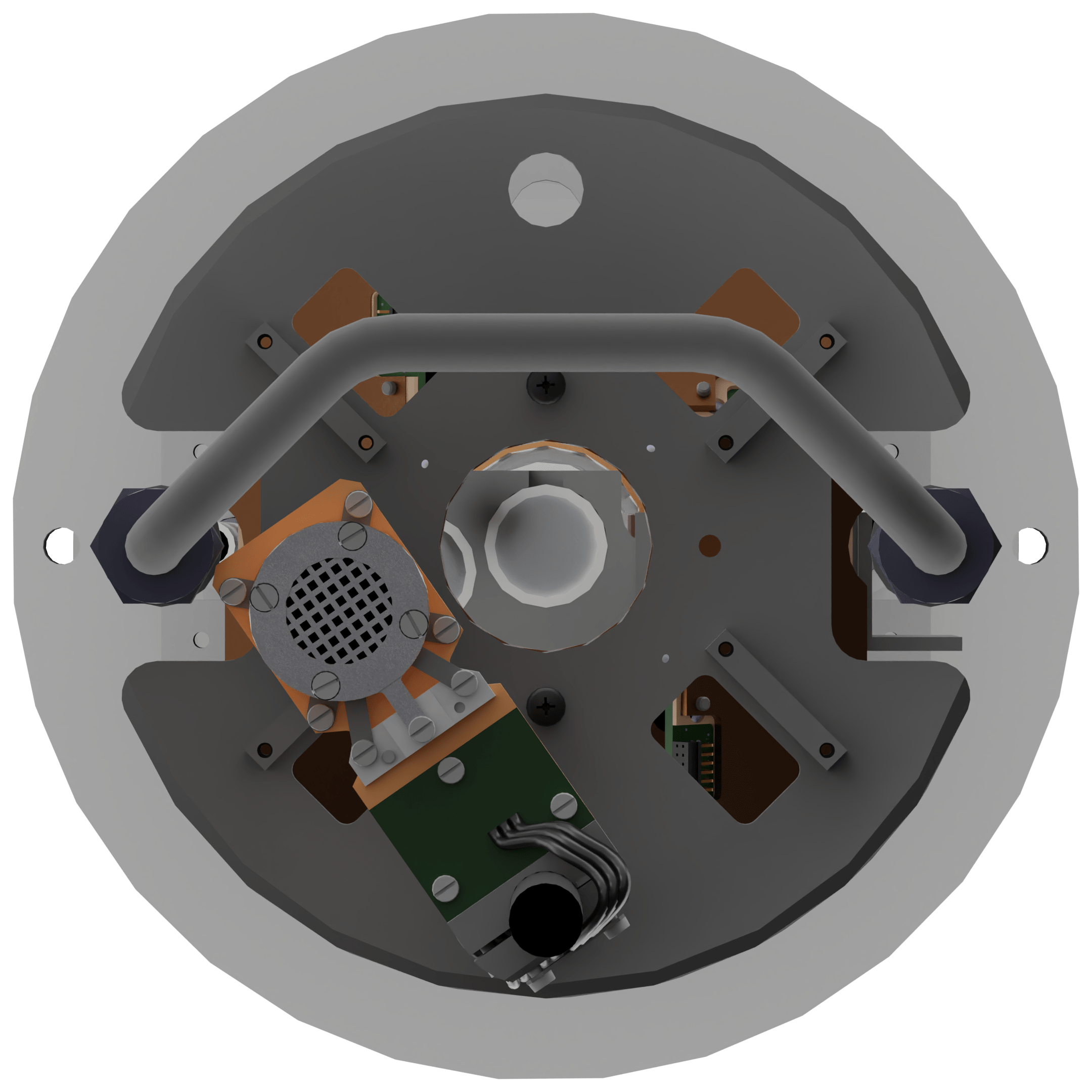}
        \centering \text{(a)}
    \end{minipage}
    \begin{minipage}[r]{0.49\linewidth}
        \includegraphics[width=1\linewidth]{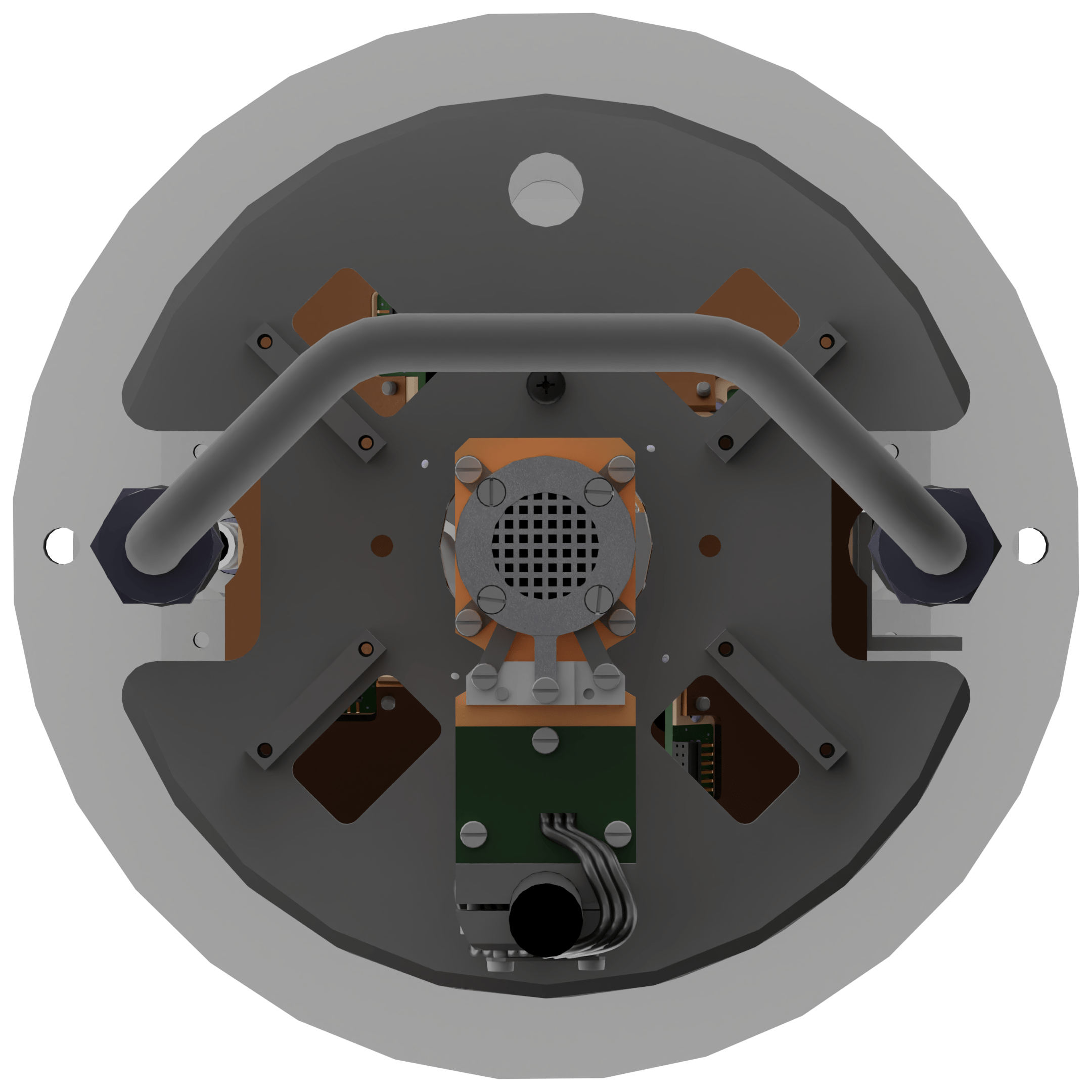}
        \centering \text{(b)}
    \end{minipage}
    \caption{Bottom view of the WISArD experimental tower with the MCP position \textit{out} (a) and \textit{in} (b). The catcher foil stopping the beam during data taking is held by the white PEEK ring at the center of the tower. The calibration mask geometry is the 2024 version.}
    \label{FIG_tower_bottom_view}
\end{figure}

\begin{figure}[ht]
	\centering 
	\includegraphics[width=0.8\columnwidth]{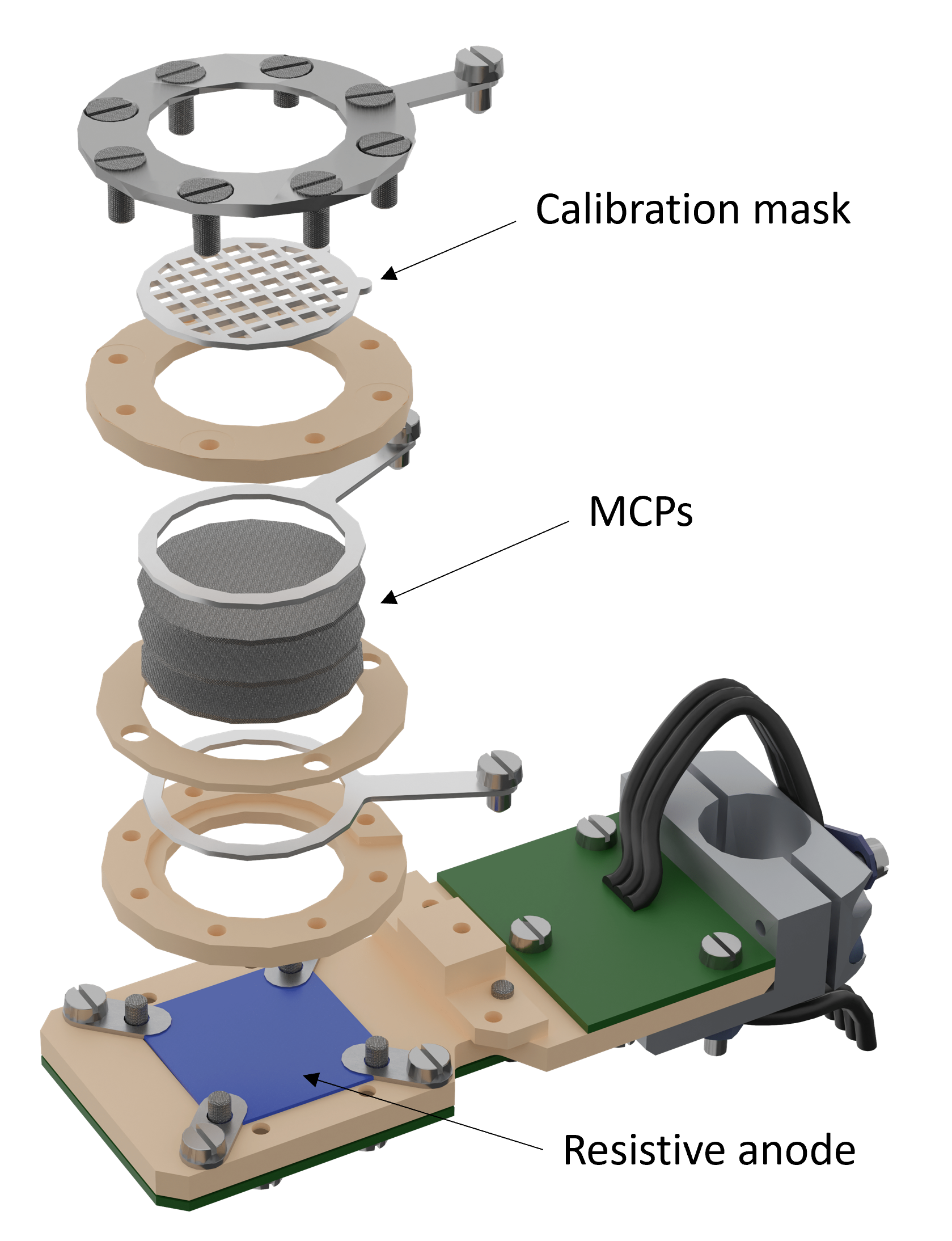}	
	\caption{Exploded view of the MCP detector assembly. See text for details.} 
	\label{FIG_MCP_Assembly}
\end{figure}

\begin{figure}[ht]
	\centering 
	\includegraphics[width=0.7\columnwidth]{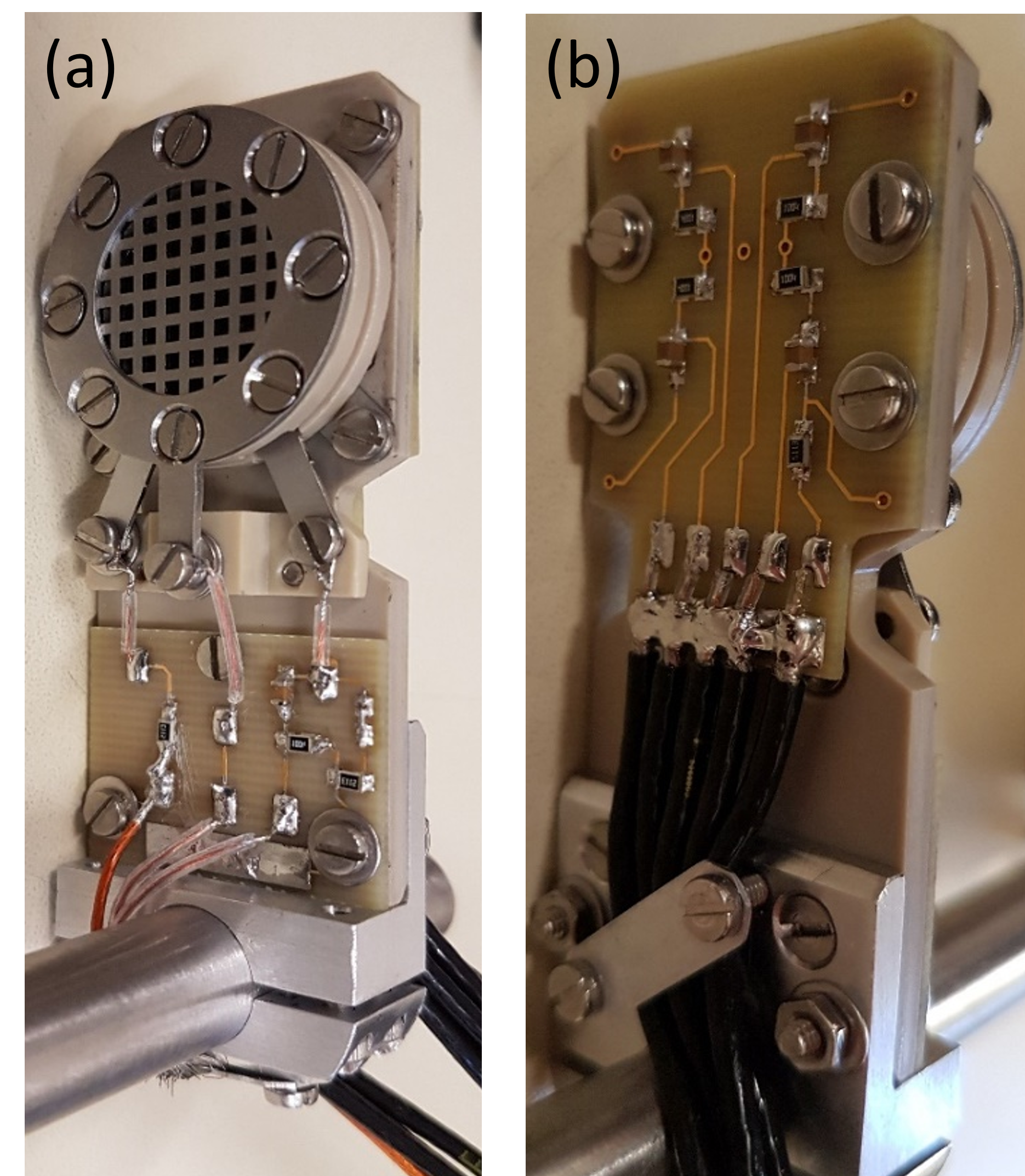}	
	\caption{Picture of the front (a) and back (b) side of the detector with the 2024 mask geoemtry.} 
	\label{FIG_MCP_picture}
\end{figure}

The MCP detector was designed to be inserted upstream of the detector tower (Fig.~\ref{FIG_tower}). It is mounted on one of the vertical rotatable rods, allowing it to be moved out of the central axis position to a side position, where it does not intercept the incoming beam or the decay products emitted from the catcher, as shown in Fig.~\ref{FIG_tower_bottom_view}. These constraints limited the detector assembly to a maximum thickness of 15~mm and a maximum width of 25~mm. The resulting compact design is shown in Fig.~\ref{FIG_MCP_Assembly}. Note that in Fig.~\ref{FIG_tower_bottom_view} and Fig.~\ref{FIG_MCP_picture} the calibration mask is the 2024 geometry having lines merging in the center, while the data uses in 2025 version with a central hole in the center.

The main holder that houses the resistive anode as well as the MCP centering rings are made of PEEK. Passive electronic components (surface-mount devices) required for fast signal decoupling and detector operation are directly mounted on two printed circuit boards (PCBs) fixed to the main PEEK holder (see Fig.~\ref{FIG_MCP_picture}). The top PCB is used for the front MCP bias voltage and to collect the back MCP fast charge signal, while the bottom PCB supplies the acceleration voltage for the anode and collects charge signals from its four corners. Connection to the four corners is ensured by four thin stainless-steel strips pressed against the anode surface. A 0.5~mm thick aluminum mask with a 2~mm pitch pattern (1.4~$\times$~1.4~mm$^2$ holes) is fixed in front of the first MCP for spatial calibration. As in Ref.~\cite{Hong_2016}, the mask remains permanently mounted on the detector to provide reliable online calibration, even in the event of small electronic gain drifts. With only 36~\% transparency, the mask can also function as a Faraday cup when tuning a stable beam with intensities in the pA range.

\subsection{Detector read-out}
\label{SUBSEC_Detector_Readout}

\begin{figure}[!h]
\centering
    \includegraphics[width=0.8\columnwidth]{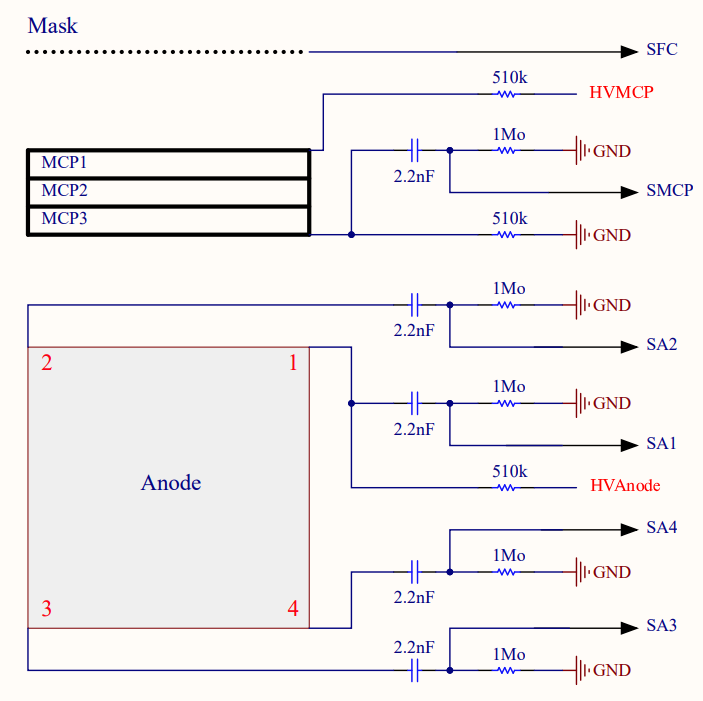}
    \caption{MCP detector electronic read-out, see text for details.}
    \label{FIG_readout}
\end{figure}

The circuit diagram of the MCP detector is shown in Fig.~\ref{FIG_readout}. Bias voltages are supplied to the MCPs and the anode through 510~kΩ resistors. Typical operating voltages for the front MCP range from $-$2.0~kV to $-$3.2~kV (HVMCP in Fig.~\ref{FIG_readout}), while the back MCP is connected to ground and the anode is supplied with $+100$~V (HVAnode in Fig.~\ref{FIG_readout}). Fast signals from the back MCP and from the four anode corners are decoupled from their DC voltage using a 2.2~nF capacitor and a 1~MΩ resistor connected to ground. The back MCP charge signal (SMCP in Fig.~\ref{FIG_readout}) and the anode signals (SAx in Fig.~\ref{FIG_readout}) are then sent through 50 Ω coaxial cables to a x10 gain fast current preamplifier and four charge preamplifiers, respectively. The charge preamplifiers, with a gain of $\sim$ 1~V/pC, are tuned for short signal decay times ($<$ 1~µs) to accommodate high counting rates. The output signal of the MCP fast preamplifier is digitized and integrated using a CARAS board from the FASTER data acquisition system (DAQ) \cite{faster}. The four anode signals from the charge preamplifiers are digitized and numerically filtered using a MOSAHR board from the same FASTER DAQ. All data was individually time-stamped for both online and offline trigger analysis with a time resolution of 8~ns.

\section{Characterization with stable beam}
\label{SEC_Experimental_results}
The MCP detector was characterized using the offline ion source of WISArD, which provides a stable 30~keV \n{39}{K}$^+$ ion beam with intensities up to a few nA. The beamline optics were optimized using a low-intensity beam (a few~pA) and a Faraday cup located at the end of the vertical WISArD beamline, downstream of the detection tower (for details see Ref.~\cite{Dinko_2023}). To prevent detector saturation and potential damage the beam intensity was reduced using a set of three 90\% attenuation grids installed at the entrance of the vertical beamline, resulting in a detector counting rate of a few $10^3$ counts per second (cps). This attenuation minimized pile-up effects and ensured safe operation. The beam was then scanned across the detector surface using x- and y-steerers, controlled by a dedicated PyEpics \cite{pyepics} program to map the full image of the calibration mask. Detector testing began with a front MCP bias voltage of $-$2.0~kV in the absence of a magnetic field (Sec.~\ref{SUBSEC_Coincidences} to~\ref{SUBSEC_Detector_performance}), followed by a performance evaluation with magnetic fields ranging from 1~T to 4~T (Sec.~\ref{SUBSEC_B_effect}). 

\subsection{Event selection}
\label{SUBSEC_Coincidences}
\begin{figure}[h]
    \includegraphics[width=\columnwidth]{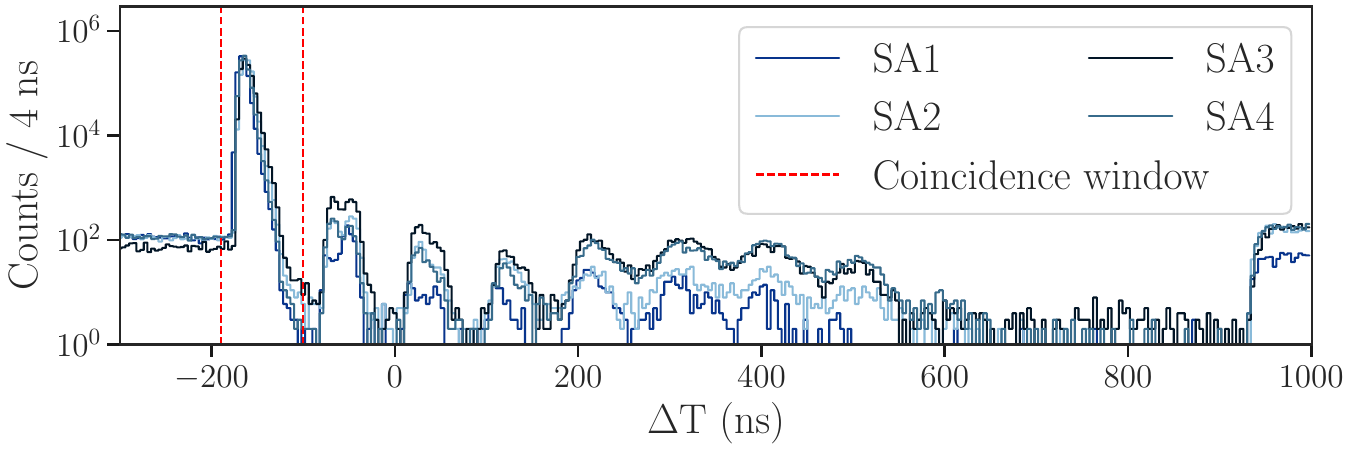}
    \caption{Time difference between the signals of each corner and the MCP signal. The limits [$-$190, $-$100]~ns of the time coincidence window used for event selection are indicated with red dashed vertical lines.}
    \label{FIG_Coinc}
\end{figure}

Figure.~\ref{FIG_Coinc} shows the time difference histograms between each of the four corner signals of the anode and the MCP signal used as a trigger. The main coicidence peak is located at $-$170~ns.
The plateau before the main coincidence peak, is caused by random coincidences. Each channel has its own intrinsic dead time, either imposed by the QDC time integration window \\(200~ns for the SMCP) or by the ADC signal sampling time (1080~ns for the SAx).
The latter leads to strong suppression of random coincidences after the main peak until $\Delta T = 950$~ns. Other peaks, occurring every 100~ns after the main one, arise from reflections in the cables. For event reconstruction, a time selection window [$-$190, $-$100]~ns is applied, indicated in Fig.~\ref{FIG_Coinc} with vertical dashed lines.
This time selection window removes events with badly reconstructed position, shown in Fig.~\ref{FIG_Scan_RAW} with red points. These represent $\sim 8\%$ of the data.

\subsection{Image reconstruction and position calibration}
\label{SUBSEC_Calibration}

With resistive anodes using the Gear-geometry, the MCP detector image reconstruction can be easily obtained by comparing the charge collected at the four corners \cite{Lampton_1979}:
\begin{equation}
  \begin{aligned}
    X_d &=& \ddfrac{-C_1+C_2+C_3-C_4}{C_{\Sigma}}     \\
    Y_d &=& \ddfrac{-C_1-C_2+C_3+C_4}{C_{\Sigma}}
    \label{EQ_charge_to_xy}
  \end{aligned}
\end{equation}

\noindent where $C_{\Sigma}$ is the sum of the four charges. $X_d$ and $Y_d$ provide relative impact coordinates associated with the four collected charges $C_i$.\\
\begin{figure}[h]
    \includegraphics[width=\columnwidth]{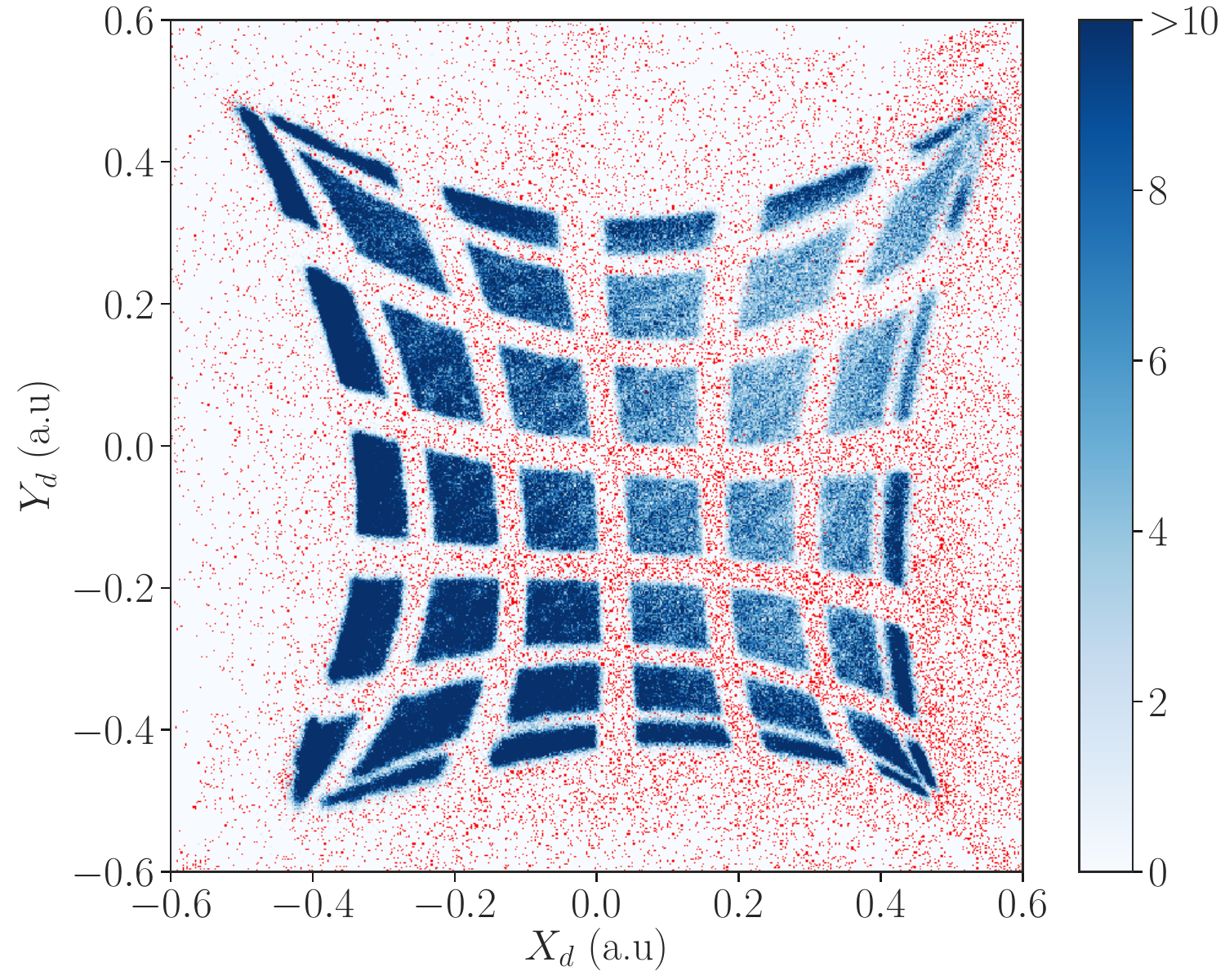}
    \caption{MCP detector scan image with stable beam reconstructed using Eq.~\eqref{EQ_charge_to_xy}. Red points represent events outside the coincidence time window (see Sec.~\ref{SUBSEC_Coincidences}).}
    \label{FIG_Scan_RAW}
\end{figure}
Figure.~\ref{FIG_Scan_RAW} shows the reconstructed image using Eq.~\eqref{EQ_charge_to_xy} from a scan of the MCP detector surface using the stable ion beam. The major pincushion distortion originates from using a square-shaped resistive anode \cite{Fraser_1981}. The image is, moreover, off-centered and the bottom right corner appears quenched. This is caused by gain differences between the four corners, particularly with the bottom right corner, which has a higher gain.\\
This distortion can be corrected using the logarithm of the normalized corner pulse heights. Keeping the same formalism as in Eq.~\eqref{EQ_charge_to_xy}, the corrected position $(X, Y)$ can then be expressed as:

\begin{equation}
  \begin{aligned}
    X &=& -\ddfrac{-c_1+c_2+c_3-c_4}{c_1+c_2+c_3+c_4}     
     \\
    Y &=& -\ddfrac{-c_1-c_2+c_3+c_4}{c_1+c_2+c_3+c_4}
    \label{EQ_logcharge_to_xy}
  \end{aligned}
\end{equation}
where $c_i = \ln(C_i/C_\Sigma)$.\\

\begin{figure}[h]
    \includegraphics[width=\columnwidth]{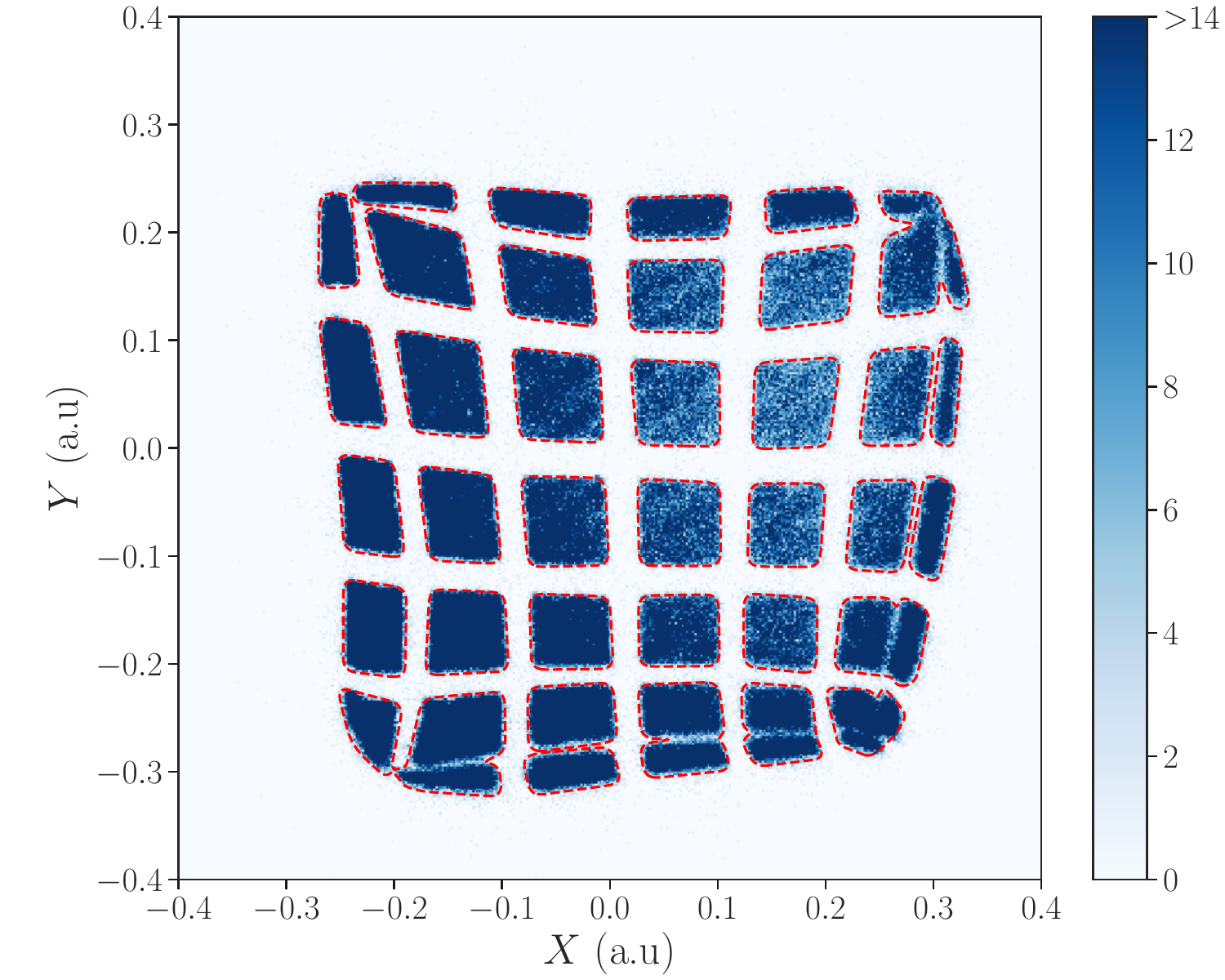}
    \caption{MCP detector scan image with the stable ion beam reconstructed using Eq.~\eqref{EQ_logcharge_to_xy} with the stable beam. The red contour is obtained for $F_{scan}(X, Y) = A/2$ (see text for details).}
    \label{FIG_Scan_fitted}
\end{figure}

\begin{figure}[!h]
\centering
    \includegraphics[width=0.9\columnwidth]{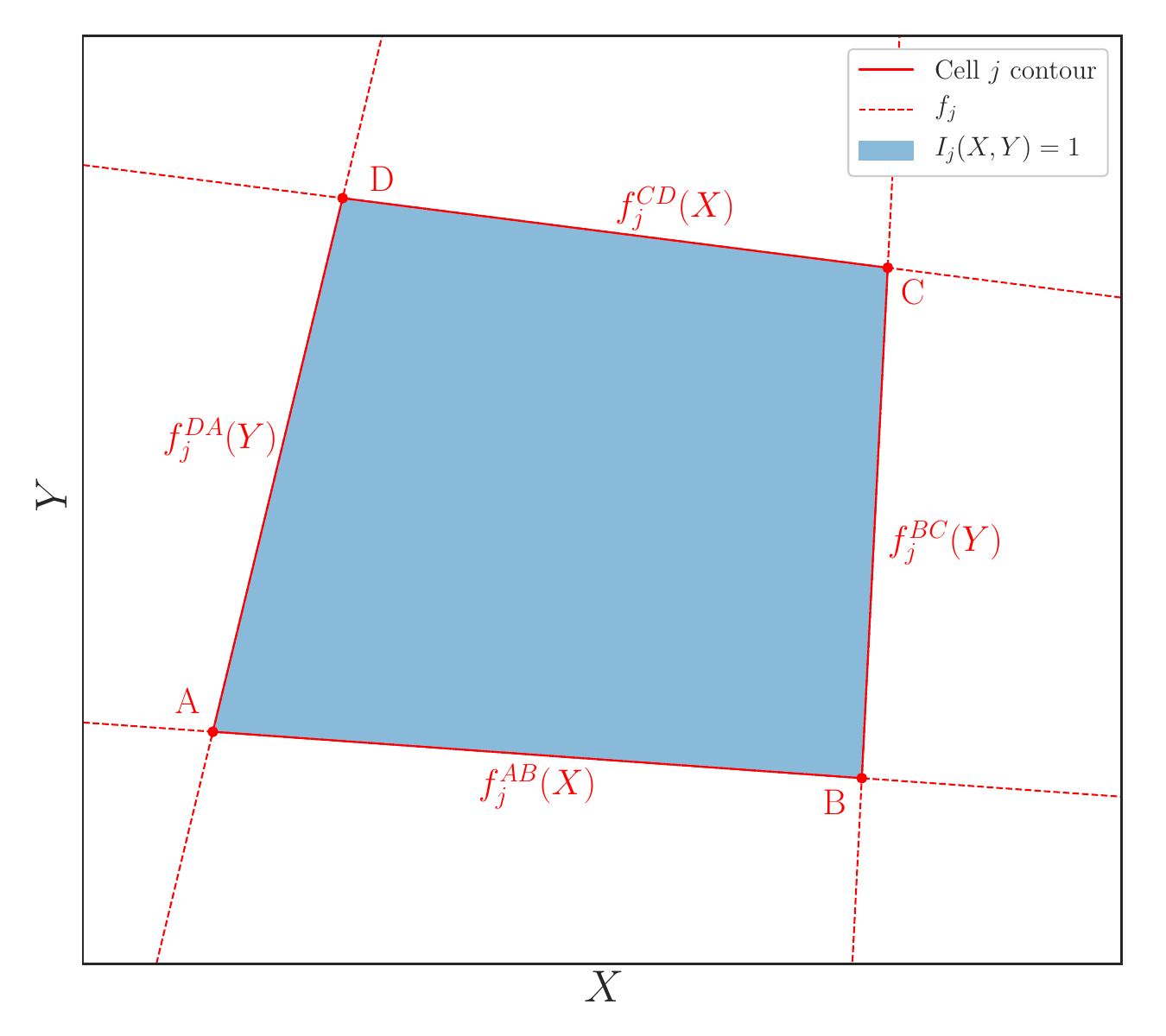}
    \caption{Graphical representation of Eq.~\eqref{EQ_Icell} modeling the calibration mask around a given hole $j$.}
    \label{FIG_GRID_SKETCH}
\end{figure}

Figure.~\ref{FIG_Scan_fitted} shows the result of the logarithmic correction using Eq.~\eqref{EQ_logcharge_to_xy}. The pincushion distortion is significantly reduced, but further corrections are required to achieve the desired precision on the ion hit position.\\
The most straightforward reference points on the reconstructed image are the corner positions of each hole of the calibration mask. These reference points were used for the detector calibration. 
The coordinates $(X_{jABCD}, Y_{jABCD})$ for each hole $j$ in Fig.~\ref{FIG_Scan_fitted} were first determined manually. 
Then, a 2D function was built to determine these coordinates more precisely using a fit of the mask image.
The contour of a deformed hole $j$ was defined by four linear functions, $f^{AB}_{j}$, $f^{BC}_{j}$, $f^{CD}_{j}$ and $f^{DA}_{j}$, derived from the coordinates of the four corners of the hole labeled $A$, $B$, $C$ and $D$, as shown in Fig.~\ref{FIG_GRID_SKETCH}:

\begin{eqnarray}
    f^{AB}_{j}(X) &= \ddfrac{Y_{jA} - Y_{jB}}{X_{jA} - X_{jB}} (X-X_{jA}) + Y_{jA} \\
    f^{BC}_{j}(Y) &= \ddfrac{X_{jB} - X_{jC}}{Y_{jB} - Y_{jC}} (Y-Y_{jB}) + X_{jB} \\
    f^{CD}_{j}(X) &= \ddfrac{Y_{jC} - Y_{jD}}{X_{jC} - X_{jD}} (X-X_{jC}) + Y_{jC} \\
    f^{DA}_{j}(Y) &= \ddfrac{X_{jD} - X_{jA}}{Y_{jD} - Y_{jA}} (Y-Y_{jD}) + X_{jD}  \,\,.
    \label{EQ_cell_contour}
\end{eqnarray}

The mask transmission can then be expressed as: 
\begin{equation}
    I_j(X, Y) = 
        \begin{cases}
            1, \text{if } f^{AB}_{j} < Y \;\land\; f^{BC}_{j} > X \;\land\; f^{CD}_{j} > Y \;\land\; f^{DA}_{j} < X \\
            0, \text{ otherwise}.
        \end{cases}
    \label{EQ_Icell}
\end{equation}

When an ion hits the detector mask material at position $(X, Y)$, it is stopped, and  $I_j(X, Y) = 0$. Conversely, if an ion passes through a hole, it is detected and the transmission function is $I_j(X, Y) = 1$.

To describe the full mask geometry for the $N$ holes, the individual transmission functions have to be summed:
\begin{equation}
    I_{mask}(X, Y) = \sum_{j=1}^{N} I_j(X, Y).
    \label{EQ_Igrid}
\end{equation}

To take the spacial resolution of the detector into account, Eq.~\eqref{EQ_Igrid} is convoluted with a two-dimensional Gaussian:
\begin{equation}
    F_{scan}(X, Y) = A \left(I_{mask} * R(\delta_X, \delta_Y)\right) (X, Y)
    \label{EQ_fit_scan}
\end{equation}

\noindent where $R$ is a 2D Gaussian centered on $(0,0)$ with standard deviations in $X$ and $Y$ defined as $\delta_X$, $\delta_Y$ respectively, and $A$ is the number of counts. 
This convolution has an analytical expression:
\begin{eqnarray}
F_{scan}(X, Y) = \frac{A}{16} \sum_{j=1}^{N} \Bigg\{
&\Bigg[1 + \operatorname{erf}\left(\frac{Y - f^{AB}_{j}(X)}{\sqrt{2}\delta_Y} \right)\Bigg]& \nonumber\\  
\times&\Bigg[1 - \operatorname{erf}\left(\frac{X - f^{BC}_{j}(Y)}{\sqrt{2}\delta_X} \right)\Bigg]& \nonumber\\
\times&\Bigg[1 - \operatorname{erf}\left(\frac{Y - f^{CD}_{j}(X)}{\sqrt{2}\delta_Y} \right)\Bigg]& \nonumber\\
\times&\Bigg[1 + \operatorname{erf}\left(\frac{X - f^{DA}_{j}(Y)}{\sqrt{2}\delta_X} \right)\Bigg]& \Bigg\}.
\label{EQ_fit_scan_final}
\end{eqnarray}

$F_{scan}(X, Y)$ was fitted to the data to extract all corner positions. A maximum number of counts was applied to the bin contents of Fig.~\ref{FIG_Scan_fitted} to impose a common amplitude $A$ for 45 holes. This reduces the number of free parameters to $2 \times 4 \times 45$ parameters for the corner positions, two for the resolution in x and y, and one for the amplitude. 
The result of the fit is shown in Fig.~\ref{FIG_Scan_fitted}, with red dotted lines connecting the corner coordinates $(X_{jABCD}, Y_{jABCD})$ that were obtained.

\begin{figure}[h]
    \includegraphics[width=\columnwidth]{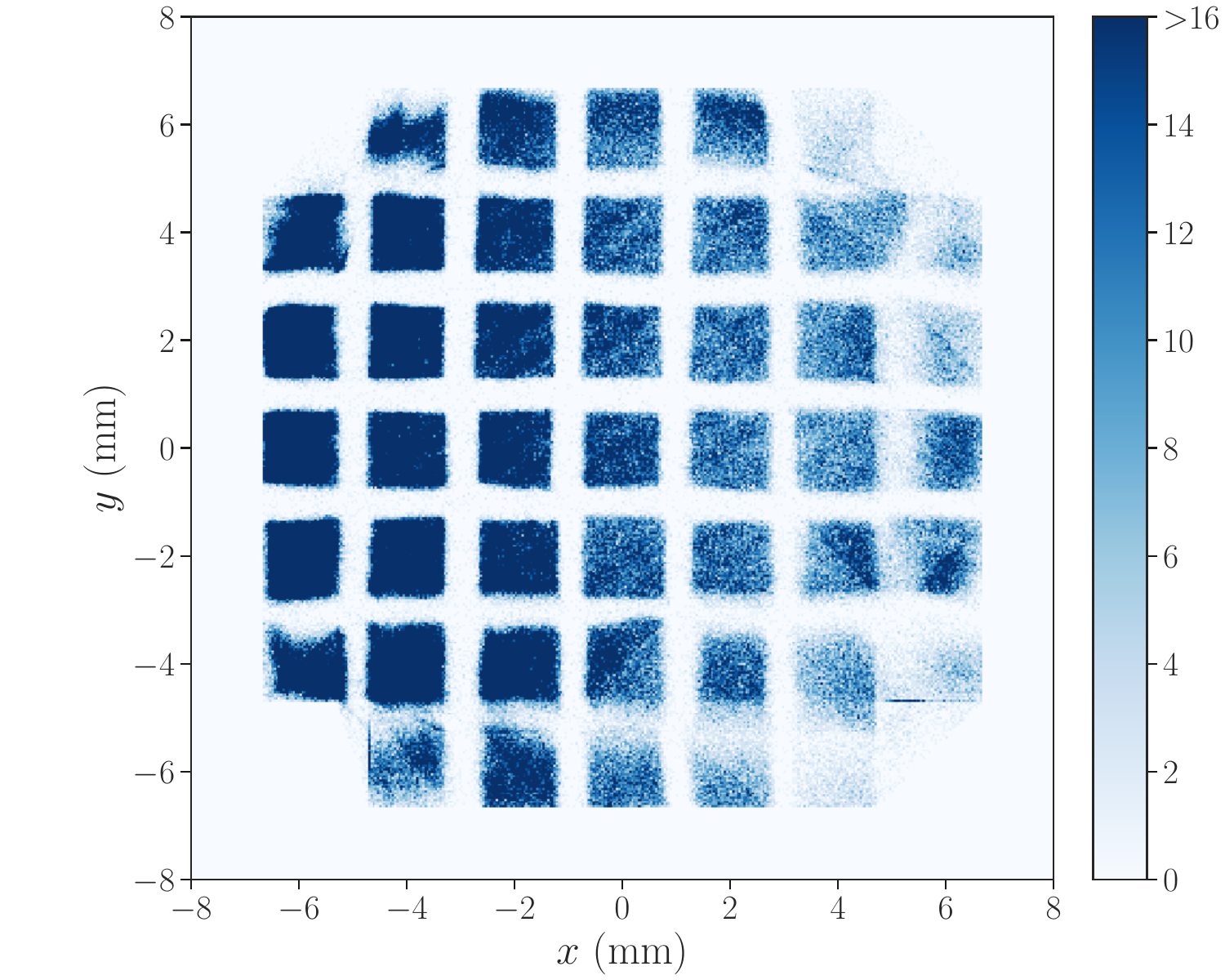}
    \caption{MCP detector scan image reconstruction after calibration and interpolation procedure.}
    \label{FIG_Scan_calibrated}
\end{figure}

The corner coordinates $(X_{jABCD}, Y_{jABCD})$ can then be used to define two bilinear interpolation functions, $x(X, Y)$ and $y(X, Y)$, mapping all the $(X_{jABCD}, Y_{jABCD})$ coordinates to their corresponding positions $(x_{jABCD}, y_{jABCD})$ on the mask frame.
When applied to the data of Fig.~\ref{FIG_Scan_fitted}, these functions now provide calibrated coordinates, resulting in the scan image shown in Fig.~\ref{FIG_Scan_calibrated}. 

\begin{figure}[h]
    \includegraphics[width=\columnwidth]{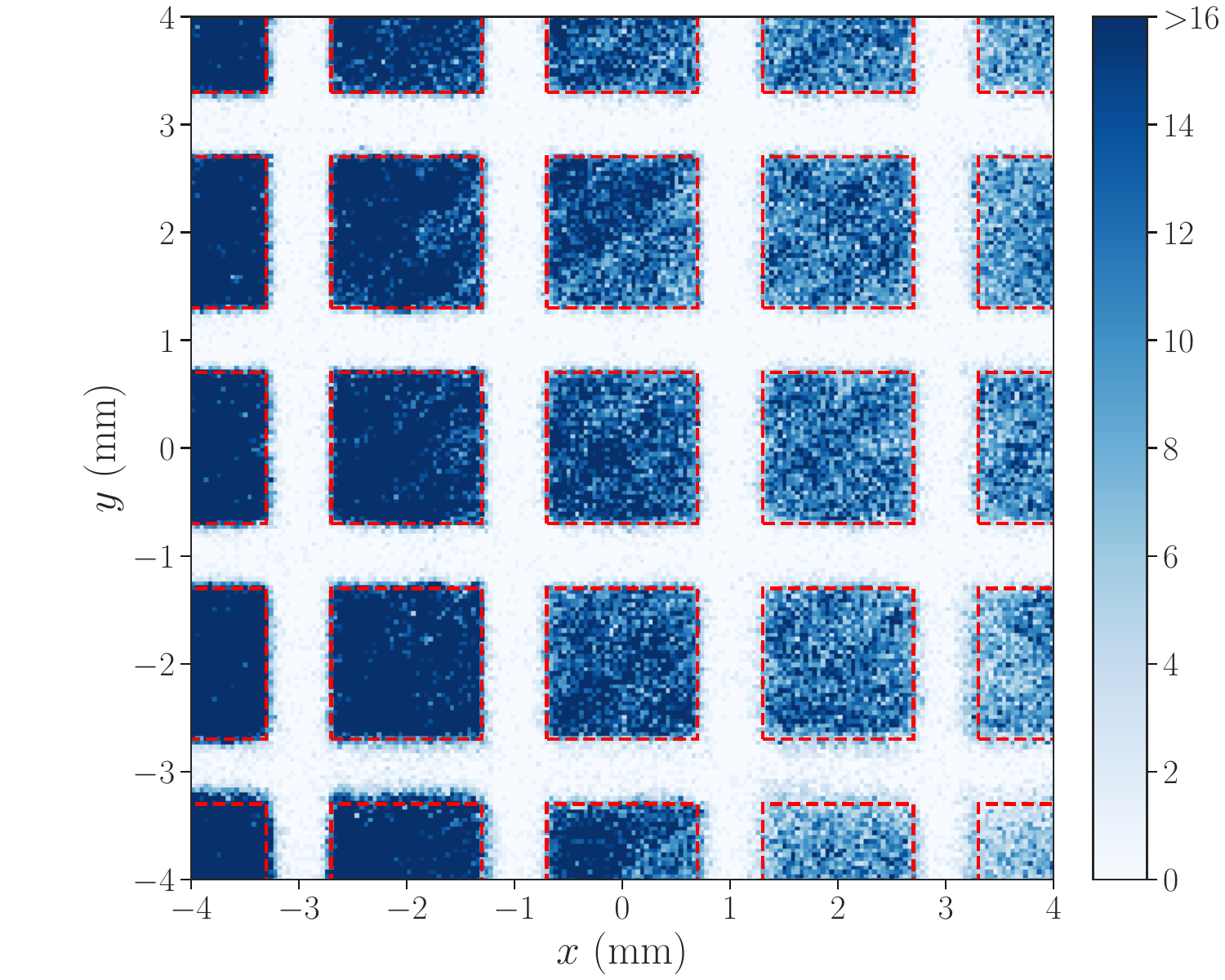}
    \caption{MCP detector scan image reconstruction after the final calibration and interpolation procedure zoomed in the ROI. Red contours represent the physical geometry of the mask grid.}
    \label{FIG_Scan_calibrated_2}
\end{figure}

To further improve the accuracy of the calibration, the fit procedure was performed again with a refined $F_{scan}(x, y)$ fit function applied to the histogram of Fig.~\ref{FIG_Scan_calibrated}. In this final calibration step, the bin content of the histogram was left untouched and the common amplitude $A$ of Eq.~\eqref{EQ_fit_scan_final} was replaced by N free independent parameters $A_j$ associated to each hole. Similarly, the resolution parameters $\delta_{x}$ and $\delta_{y}$ were each replaced by N independent parameters $\delta_{xj}$ and $\delta_{yj}$. The N hole images were then fitted independently to limit the number of free parameters and facilitate the fit convergence. A final bilinear interpolation provides refined $x$ and $y$ 
coordinates leading to the reconstructed image in Fig.~\ref{FIG_Scan_calibrated_2}. In the detector region of interest (ROI), the data now overlap almost perfectly with the mask geometry shown with dotted red lines.

\subsection{Detector resolution and performances}
\label{SUBSEC_Detector_performance}

\begin{figure}[h]
    \centering
        \includegraphics[width=1.0\columnwidth]{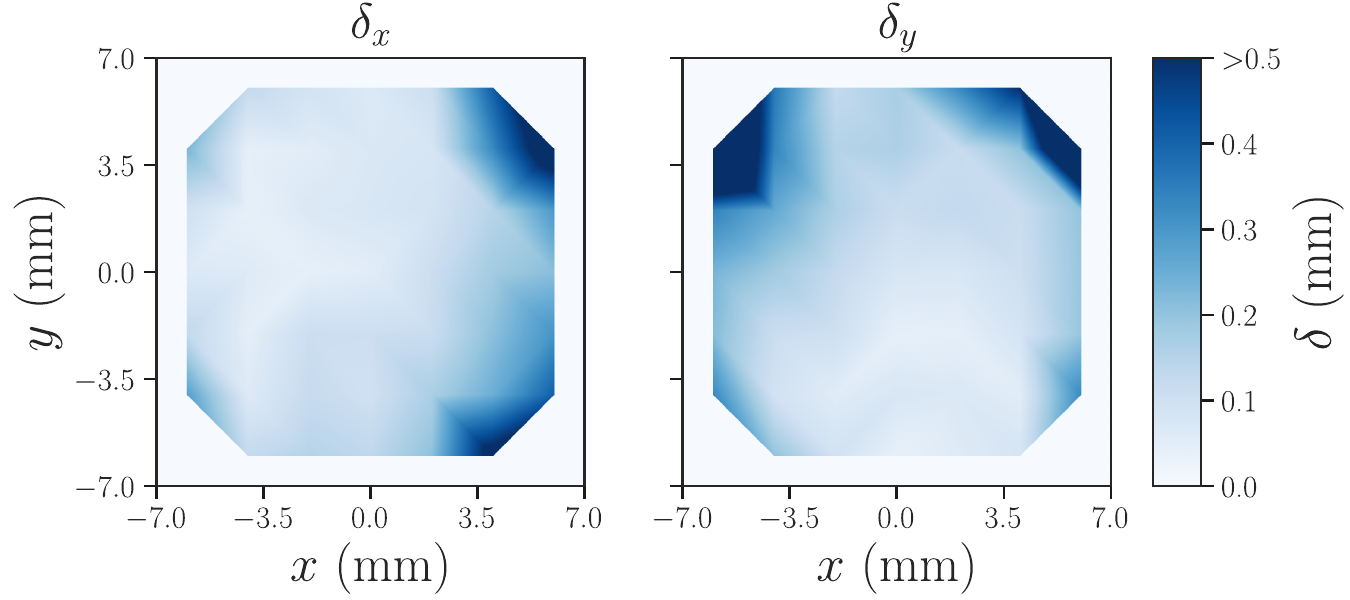}
        \caption{Bilinear interpolation between fitted resolution $\delta_{xj}$ and $\delta_{yj}$}
        \label{FIG_Resolution}
\end{figure}

The $\delta_{xj}$, $\delta_{yj}$ values resulting from the final fit were interpolated as functions of the coordinates (x,y) in the mask frame and are shown in Fig.~\ref{FIG_Resolution}. For the larger part of the mask surface, the resolutions $\delta_{x}$ and $\delta_{y}$ are better than 0.1~mm. A degradation is observed close to the corners, where the initial deformations are larger (Fig.~\ref{FIG_Scan_fitted}) and the signal to noise ratio is lower.

Within the ROI, the detector resolution is also characterized using the final reconstructed image of Fig.~\ref{FIG_Scan_calibrated_2}, projections of the 9 central holes were made on the $x$ and $y$ axes. These projections were fitted with an error function at the horizontal and vertical edges, providing for each hole two $\delta_{xj}$ and two $\delta_{yj}$ values. The mean values and standard deviations of the 36 fit results obtained in both x and y dimensions, $\overline{\delta}_{x} = 0.061(3)~\text{mm}$ and $\overline{\delta}_{y} = 0.063(5)~\text{mm}$, confirm the good resolution and homogeneity of the detector within the ROI.

\subsection{Effect of the magnetic field}
\label{SUBSEC_B_effect}

\begin{figure}[!h]
    \centering
    \includegraphics[width=\columnwidth]{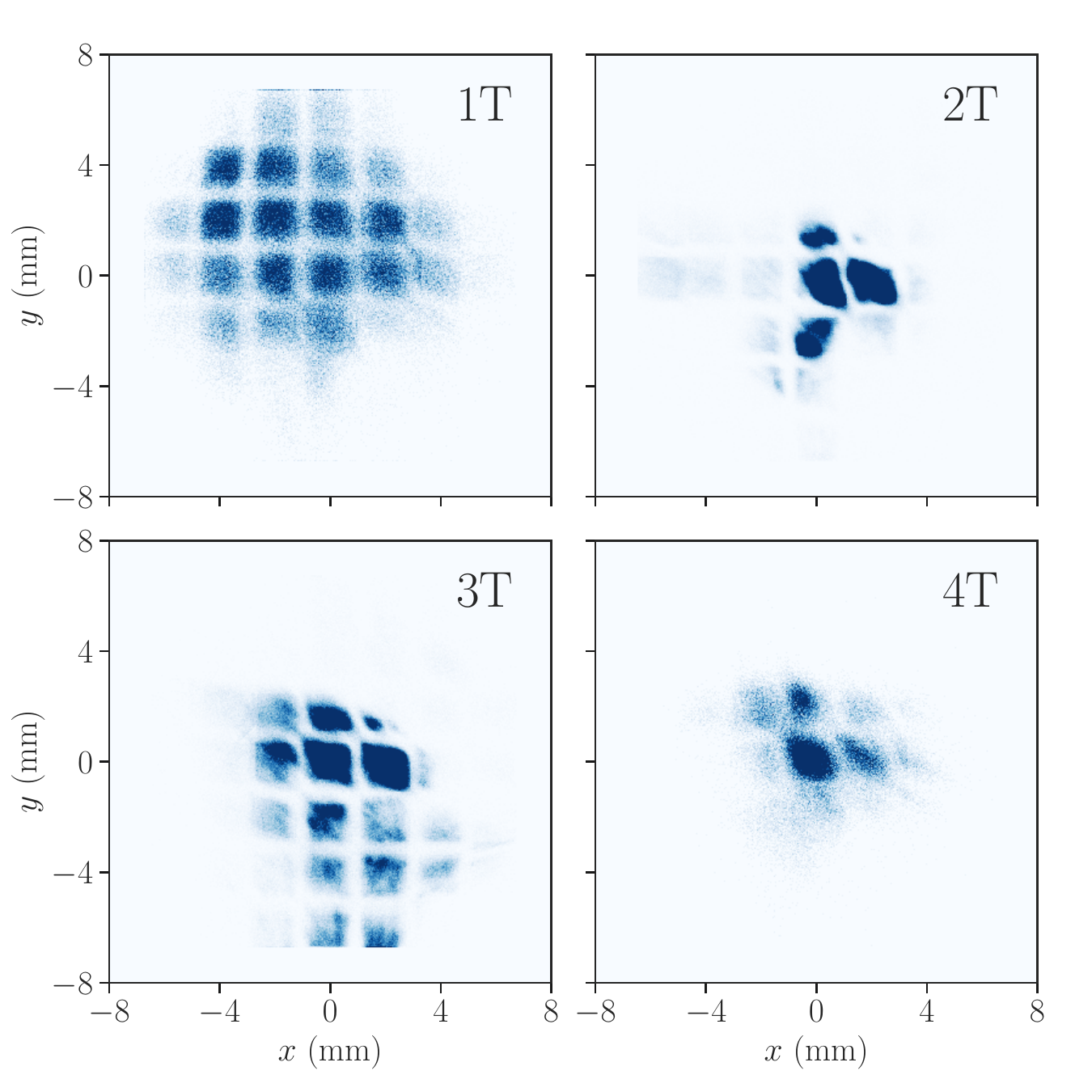}
    \caption{MCP detector scan image reconstruction using the bilinear interpolation used for the Fig.~\ref{FIG_Scan_calibrated_2} and the same steerer sweeping range. The bias voltages used are summarised in Table~\ref{tab:res} with $-$2.0~kV at 1T.}
    \label{FIG_Fields_reconstruction}
\end{figure}

The MCP detector calibration described in Sec.~\ref{SUBSEC_Calibration} was performed without magnetic field and with a detector bias of $-$2.0~kV. Ideally, with the permanent mask in front of the MCPs, the same method could be applied up to 4~T, which is the nominal magnetic field value for radioactive beam measurements. However, such calibrations were not possible, as the ion beam confinement induced by the strong magnetic field prevented us from scanning the beam over the full mask surface with a reasonable homogeneity. 
As shown in Fig.~\ref{FIG_Fields_reconstruction}, for magnetic field above 1~T, the intense part of the ion beam could reach only specific regions of the mask, covering only partially a few holes, and with a very large contrast between the areas hit by the main beam (intense blue) and ions with more marginal trajectories (light blue). 
The images of beam scans obtained with magnetic fields ranging from 1~T to 4~T and shown in Fig.~\ref{FIG_Fields_reconstruction} were thus reconstructed using the calibration parameters obtained without magnetic field and analyzed using the hole projection method Sec.~\ref{SUBSEC_Detector_performance}. 
When increasing the magnetic field, the gain of the MCP decreases significantly, which leads to a loss of both efficiency and spatial resolution. For each field value, a high-voltage threshold was observed below which no signal could be detected. This minimum voltage can be roughly estimated as HVMCP$_{min} \simeq -1.8 - 0.2\times \text{B}$ (with HVMCP$_{min}$ in kV and B in T). 
To mitigate the effect on detection efficiency and resolution, the high voltage was increased, up to a limit of $-$3.2~kV, which was chosen to avoid damaging the MCPs when operating at 4~T. Table~\ref{tab:res} displays the mean resolution $\overline{\delta}_{x}$ and $\overline{\delta}_{y}$ obtained for various combination of magnetic field and high-voltage. 
Note that, because of the large inhomogeneity in the beam scan, only a limited number of vertical $n_y$ and horizontal $n_x$ edge projections could be fitted to provide an estimate of the resolution. The poor quality of the 4~T scan did not allow us to extract reliable data.
At $-$2.0~kV, the resolution was degraded by a factor ~3.3 when increasing the magnetic field from 0~T to 1~T. Resolutions of the order of 0.15~mm were typically obtained between 1 and 3~T with the bias voltages that were tested, showing a significant degradation compared to the results obtained without magnetic field. 
\begin{table}[!htb]
    \caption{\label{tab:res} Spatial resolution for different magnetic field values and bias voltages. The error corresponds to the standard deviation of the measurements and n$_{x,y}$ indicates the number of measurements. }   
    \vspace{1 mm}
    \centering
    \begin{tabular}{lccccc}
        \hline\hline
        \\[-1em]
        B (T) & HVMCP (kV) & n$_x$  &  n$_y$ & $\overline{\delta}_{x}$ (mm) & $\overline{\delta}_{y}$ (mm)\\ \hline
        0.0 & $-$2.0 & 18 & 18 & 0.061(3) & 0.063(5) \\
        1.0 & $-$2.0 & 22 & 15 & 0.202(9) & 0.210(9) \\
        1.0 & $-$2.2 & 7 & 6 & 0.133(4) & 0.118(4) \\
        2.0 & $-$2.4 & 6 & 4 & 0.149(10) & 0.151(17) \\
        3.0 & $-$2.6 & 10 & 9 & 0.160(10) & 0.153(10) \\
        4.0 & $-$3.2 & 0 & 0 & - & - \\ \hline
    \end{tabular} 
\end{table}

Although the magnetic field significantly affects the gain of the MCPs, and thus the resolution, it is not expected to have a sizable effect on the charge fraction collected at the four corners of the resistive anode. This is confirmed by the images displayed in Figures~\ref{FIG_Scan_calibrated_2} and~\ref{FIG_Fields_reconstruction}, reconstructed with the same calibration parameters and showing no visible displacement of the hole pattern when increasing the magnetic field.
The scan image obtained at 1~T and $-$2.0~kV covers a sufficient number of holes with acceptable inhomogeneities to extract reliable values of the horizontal and vertical edge positions when fitting the hole projections with the resolution as a free parameter.
For a more quantitative study of the possible modifications in the image reconstruction due to the presence of the magnetic field, the data is then compared to the nominal mask positions. 
With 11 measurements on the $x$-axis and 10 on the $y$-axis within the ROI, we obtained the mean deviations $\overline{\Delta}_{x}=-0.012(83)$~mm and $\overline{\Delta}_{y}=-0.039(71)$~mm. The error corresponds to the standard deviation of the results and is mostly due to inhomogeneities in the exposure of the hole surfaces.

In the future, dedicated measurements with an alpha source should be performed to characterize the detector response at different high voltages and magnetic fields with higher precision. As the high-energy alpha particles being weakly affected by the magnetic field, a complete and homogeneous image of the mask can then be obtained up to 4~T.

This would allow a systematic study of the resolution, gain stability, and position reconstruction accuracy across the full parameter space.

\section{\n{32}{Ar} beam profile measurement}
\label{SEC_Beam_Profile}
The radioactive beam of \n{32}{Ar} from ISOLDE is produced by bombarding a nano-structured CaO target with 1.4~GeV protons and is delivered to the experiment using the High-Resolution Separator (HRS) set to mass 32 for singly-charged atoms. During the beam profile measurements, the magnetic field was set to 4~T, and the vacuum in the detection chamber was maintained at approximately $10^{-7}\,\text{mbar}$.

\subsection{Beam profile extraction}
\label{SUBSEC_Beam_Profile}

\begin{figure}[h]
    \includegraphics[width=\columnwidth]{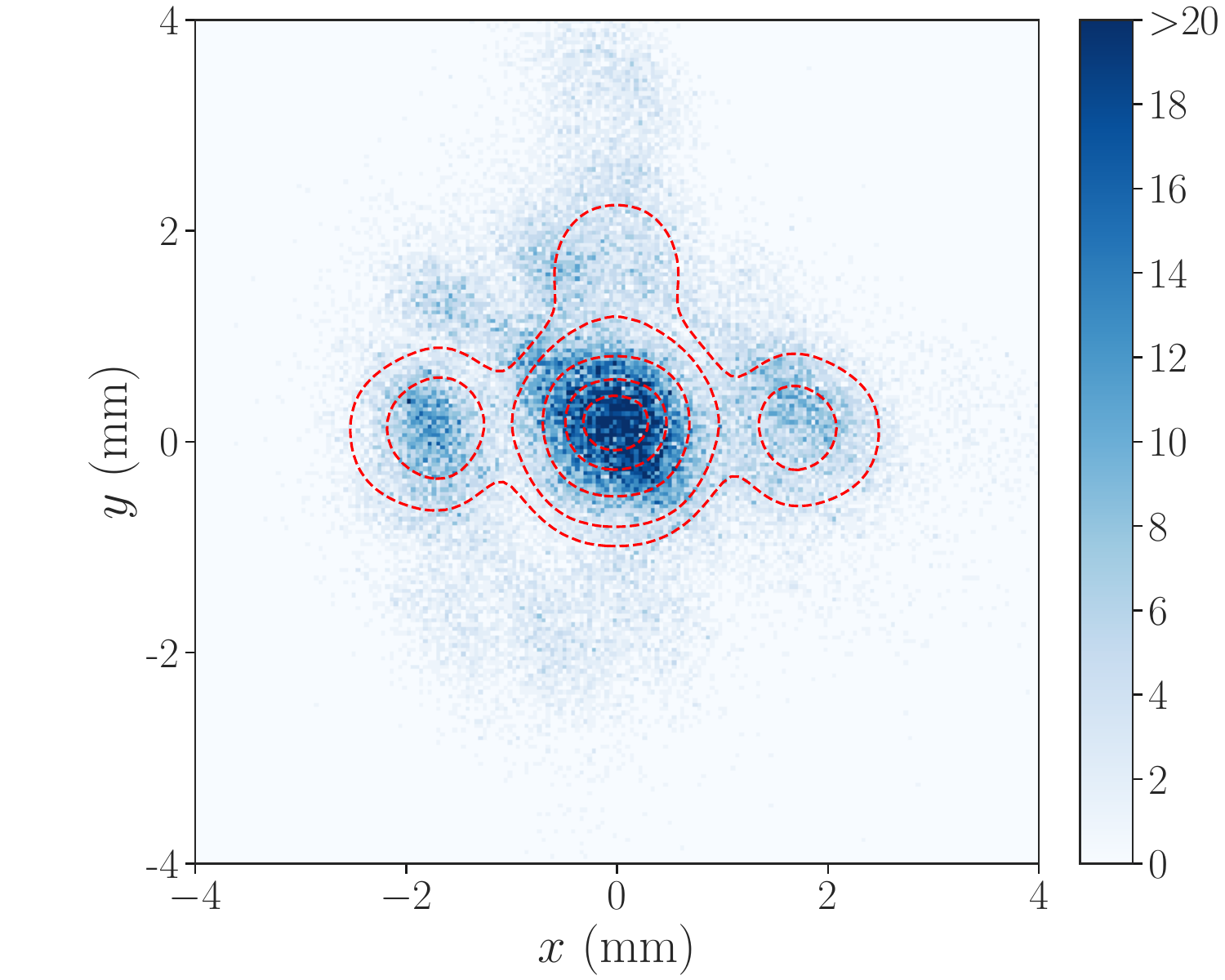}
    \caption{Fitted MCP image of the radioactive beam of $^{32}\mathrm{Ar}$ with the 2025 mask design at 4~T and $-$3.2~kV zoomed in the ROI. The red contours represent 15\%, 25\%, 50\%, 75\% and 90\% of $max(F_{beam}(x, y))$.}
    \label{FIG_Beam_fitted}
\end{figure}

After characterizing the MCP detector behavior in magnetic fields, we use this knowledge to obtain the \n{32}{Ar} beam profile, needed for a precise determination of $\tilde{a}_{\beta\nu}$.
The beam profile was modelled as a 2D Gaussian distribution $G_{beam}(A_{beam}, \mu_x, \mu_y, \sigma_x, \sigma_y)$ centered at $(\mu_x, \mu_y)$ with standard deviations $(\sigma_x, \sigma_y)$ and an amplitude $A_{beam}$. Replacing in Eq.~\eqref{EQ_fit_scan} the constant $A$ by the 2D Gaussian distribution, the beam image in the mask frame can be expressed as:

\begin{align}
   & F_{beam}(x, y) =\nonumber\\  
 &\Bigg(\bigg[I_{mask} G_{beam}(A_{beam}, \mu_x, \mu_y, \sigma_x, \sigma_y) \bigg]* R(\delta_x, \delta_y) \Bigg)(x, y)
    \label{EQ_beam}
\end{align}

\noindent where $G_{beam}$ is the 2D Gaussian function. $I_{mask}$ and $R$ are defined in Eqs.~\eqref{EQ_Igrid} and \eqref{EQ_fit_scan}, replacing the $X$ and $Y$ coordinates by $x$ and $y$ and using the mask corner coordinates $(x_{jABCD}, y_{jABCD})$.
The beam image shown in Fig.~\ref{FIG_Beam_fitted} was reconstructed using the calibration parameters determined at 0~T and fitted with the 
$F_{beam}(x, y)$ function to extract the beam shape parameters, $(\mu_x, \mu_y)$ and $(\sigma_x, \sigma_y)$.
As no analytical solution for the current form of $F_{beam}(x, y)$ exists, a numerical convolution has been implemented.

The fit results indicate a beam spot centered at \\$\mu_x$ = $-0.05(8)_{stat}$~mm and $\mu_y = 0.41(6)_{stat}$~mm, with widths of $\sigma_x = 1.05(7)_{stat}$~mm and $\sigma_y = 0.61(8)_{stat}$~mm. 
The resolution of the MCP detector was considered constant within the ROI and involved two additional free parameters, $\delta_{x}$ and $\delta_{y}$, with the resulting fitted values of $0.33(4)_{stat}$~mm and $0.39(4)_{stat}$~mm, respectively.
Figure.~\ref{FIG_Beam_fitted} shows the reconstructed image of the \n{32}{Ar} beam profile, overlaid with isocontours of the fit function. The fit reproduces the beam shape rather accurately and the assumption of a 2D Gaussian beam shape provides a reasonable first-order description of the beam profile. 

This hypothesis could be refined using a dedicated \- ion-transport simulation. In particular, a SIMION simulation incorporating the WISArD beam-line optics and the applied magnetic field would provide a more realistic estimate of the beam distribution at the catcher plane. Such simulations could help quantify potential deviations from the Gaussian shape.
\section{Summary and conclusions}
Integration of beam diagnostics into the WISArD experimental setup presents significant challenges due to both mechanical constraints and the 4 T magnetic field environment. These constraints led to the selection of a compact MCP-based detector coupled to a square-shaped resistive anode. The MCP gain reduction due to the high magnetic field was mitigated by the use of three MCPs in a Z-stack configuration with small pore diameter and bias angle. 
Although the square-shaped anode introduces geometrical distortions, calibration and correction procedures enabled beam profile reconstruction with sub\-millimeter-level accuracy. During the most recent WISArD experimental campaign, the detector successfully measured the \n{32}{Ar}$^+$ beam profile under nominal 4~T conditions.
The fitted beam parameters were used to propagate the uncertainties in the beam profile to the observable of interest $\tilde{a}_{\beta\nu}$ using Geant4 simualtions. This procedure will be described in detail in a future paper on the WISArD analysis. The result, $(\Delta\tilde{a}_{\beta\nu})_{beam} = 0.7(1)$\textperthousand, was found to be compatible with the final precision aimed at by the WISArD experiment.


\section*{Acknowledgments}
This project has received funding from the Horizon Europe Research and Innovation program from the European Union under Grant Agreement No 101057511. This work was partly funded by the ANR grant ANR-18-CE31-000402 and the International Research Infrastructures grants I002619N and \\I001323N of the FWO Research Foundation-Flanders. J.Ha acknowledges support by the Institute for Basic Science (IBS-R031-Y3). D.Zakoucky acknowledges support by the Ministry of Education, Youth and Sports of the Czech Republic project LM2023040.



\bibliographystyle{elsarticle-num} 
\bibliography{example}





\end{document}